\documentclass[aps,prb,twocolumn,epsfig]{revtex4}  
\usepackage{epsfig}

\newcommand{\cugeo}{$\rm CuGeO_3$ }
\newcommand{\borate}{ $\rm SrCu_2(BO_3)_2$ }
\newcommand{\navo}{ $\rm NaV_2O_5$ }
\newcommand{\DM}{Dzyaloshinski-Moriya }

\begin{document}

\title{Theory of phonon-assisted ``forbidden'' optical transitions in spin-gapped systems}

\author{O. C\'epas$^{a,b}$, and T. Ziman$^b$}

\affiliation{
a. Department of Physics, Indian Institute of Science, Bangalore 560012, India. \\
b. Institut Laue Langevin, BP 156, F-38042 Grenoble cedex
9, France.}

\date{\today}

\begin{abstract}
We consider the absorption of light with emission of one $S_{tot}=1$
magnetic excitation in systems with a spin gap induced by quantum
fluctuations.  We argue  that an electric dipole transition is
allowed on the condition that a virtual phonon instantaneously breaks
the inversion symmetry. We derive an effective operator for
the transition and  argue that the proposed theory explains the
polarized experiments in \cugeo and \borate.
\end{abstract}

\pacs{PACS numbers: } 

\maketitle

\section{Introduction}
\label{Introduction}

Techniques of using the interactions between light and spin-waves to
study the excitations of magnetic solids were developed shortly after
the invention of the laser. Single magnon scattering of photons was
first predicted from the Zeeman coupling of the magnetic field of the
photon field to the magnetic spins, leading to magnetic dipole
transitions.\cite{BassKaganov}  Later it was pointed out
\cite{ElliottLoudon,ShenBloembergen} that the electric field of the
electromagnetic radiation could also couple to the spin, by an
indirect process in which spin-orbit interactions act on electronic
states excited virtually by electric-dipole transitions.  Experiments
in antiferromagnets \cite{FleuryLoudon} showed that this latter
mechanism dominated the magnetic-dipole transitions to single magnon
excitations. The Raman spectrum also revealed relatively strong
two-magnon scattering. This was argued \cite{FleuryLoudon} to be due
to an independent mechanism: excited-state exchange interactions.  The
same mechanism, by which the magnetic exchange interaction is modified
by electric-dipole excitation of the magnetic electrons, was advanced
\cite{Tanabe} to explain far-infrared absorption.  A variant is to
replace the virtual electronic excitation by a virtual lattice
distortion that modifies the magnetic exchange.\cite{Lorenzana} The
intensities of such transitions can be calculated by writing effective
operators for absorption or Raman scattering in terms of the spin
operators.\cite{FleuryLoudon,ElliottThorpe} This theory is considered
generally to give good account of inelastic light scattering and
optical absorption. For an isotropic system the effective operator
conserves total spin and what is commonly called the ``Fleury-Loudon''
theory is used to analyse the spectroscopy of spin conserving
transitions.

Optical techniques are now well established as probes of magnetic
excitations, whether it be by Raman scattering, i.e. inelastic
scattering of optical frequencies, electron spin resonance (ESR), i.e.
resonant absorption of electromagnetic radiation with sweeping
magnetic field, or by transmission measurements of infrared radiation.
The techniques have been further enhanced by the increasing
flexibility of light sources and detectors in the far-infrared region
that is useful to much of magnetism.  ESR studies using sources
derived from far-infrared lasers rather than the traditional cavities
are now available up to THz frequencies and may be made in large
static or pulsed magnetic fields.\cite{Motokawa} Transmission studies
in the far infrared range have the advantage of allowing for
measurement in zero external magnetic field. While restricted to small
momentum transfer, $q \approx 0$, compared to neutron inelastic
scattering, the optical techniques have the advantage of much higher
frequency resolution.  The possibility of polarising the
electromagnetic radiation means different transition mechanisms may be
distinguished.

Optical measurements are particularly useful for precise measurements
of the spin gap properties in strongly correlated systems and
spin-liquid systems with magnetic singlet ground states.  Because of
the frequencies now available, one can apply an electromagnetic source
with sufficient energy to excite the first triplet $S_{tot}=1$ excited
state from the singlet $S_{tot}=0$ ground state.
Many systems of interest are highly isotropic with
respect to spin rotations and transitions between the singlet
$S_{tot}=0$ ground state of the spin-liquid to the first triplet
$S_{tot}=1$ excited state would be forbidden by symmetry in the
isotropic limit. Even the weaker magnetic-dipole coupling should give
zero intensity as the ground state is a spin singlet.  One would then
expect to see the excited singlets, i.e. two magnon states only.
Nonetheless the ``forbidden'' transitions to the single magnon states
have been observed in many spin-liquid, ranging from the S=1/2
quasi-one-dimensional systems \cugeo
\cite{Boucher,Loosdrecht,Damascelli,Nojiri,Takehana,Room-cugeo} and
\navo \cite{NojiriNaV2O5,Takehana2}, to 2d system such as \borate
\cite{Nojiri2,Room} and to the spin-1 chain compound, NENP
\cite{Renard}.  Despite detailed experiments, no clear understanding
of the mechanism of these transitions has emerged.  It is clear that
spin-orbit coupling, that breaks the conservation of total spin, must
be included as it is then possible \textit{a priori} to have a
transition to a one-magnon state.  As mentioned, the photon can couple
to the spin degrees of freedom in different ways, via direct magnetic
dipole transitions or indirect electric dipole transitions with
spin-phonon or spin-orbit couplings.  As one of the purposes of
performing high resolution spectroscopy is to resolve the weak
anisotropies, it is important to distinguish between these mechanisms,
i.e. to find the one which gives the strongest absorption.  As in the
original studies \cite{FleuryLoudon} this is done by establishing, and
then verifying experimentally, selection rules.  For one-magnon
absorption, previous estimations favored a purely electric dipole
transition
for  NENP.\cite{Mitra} In the case of $\rm CuGeO_3$ the
suggestion that a staggered field would give rise to a magnetic dipole
transition \cite{UhrigLett} has been ruled out by the polarized
experiments.\cite{Damascelli} Furthermore the first order corrections
to the Hamiltonian in spin-orbit coupling lead to vanishing magnetic
dipole intensity owing to a lattice selection rule.\cite{Sakai} In the
compound \borate it has been shown experimentally that varying the
direction of the electric field of the wave (while keeping the
magnetic field of the wave fixed) changes the intensity of the
absorption, suggesting that the transition is electric-dipole in
nature.\cite{Room} One would also like to know which of the two
electric dipole mechanisms applies, absorption involving solely the
electronic degrees of freedom or with the lattice degrees of
freedom. In the original theory of Elliott and Loudon of light
scattering by magnons, the electric dipole coupling indeed leads to
the creation of one-magnon
excitations.\cite{ElliottLoudon,FleuryLoudon} Although such two photon
processes are not forbidden in infrared absorption, they are much
smaller in intensity since they involve the weak coupling to light to
second order in perturbation theory. Alternatively in the presence of
strong spin-orbit coupling, it is possible to have single photon
coupling to spin excitations\cite{ElliottThorpe,Rado} but as this is
of second order in the spin-orbit coupling, we shall assume that the
linear order will dominate for these materials, which are close to
isotropic.  In addition lattice symmetries such as centers of
inversion between the magnetic ion may eliminate such terms, or at
least reduce them further, if the inversion symmetry is slightly
broken, as in \borate.\cite{noteborate}

In this paper we shall  show that an effective operator of  \DM symmetry
\cite{Dzyaloshinski,Moriya} acting on the  spin degrees of freedom,

\begin{equation}
H_E = \sum_{i,a,\beta,\gamma} \textbf{\mbox{E}}^{\beta}(t) \textbf{\mbox{A}}_{\beta \gamma}(a) (\textbf{\mbox{S}}_i \times \textbf{\mbox{S}}_{i+a})^{\gamma}
\label{effectiveoperator}
\end{equation}
can be used to explain the polarized experiments of \cugeo and
\borate.  Here $\textbf{\mbox{E}}^{\beta}(t)$ is the component $\beta$
of the applied electromagnetic field at time \textit{t}. The indices
\textit{i} and \textit{a} define the lattice of magnetic bonds and the
coefficients $\textbf{\mbox{A}}_{\beta \gamma}$ will be made explicit
in section \ref{effective operator}. They couple the component $\beta$
of the electric field with the component $\gamma$ of the vector
product of the spin operators.  An electric dipole operator (1) can
arise from an electronic mechanism, as may be the case in NENP
\cite{Mitra}, but centers of inversion at the middle of the Cu-Cu
bonds in \cugeo and \borate\cite{noteborate}, would forbid generation
of the operator from purely electronic processes.  A lattice
distortion may, however, break the inversion symmetry
\textit{instantaneously}, and allow terms of the form in (1). We
therefore consider the phonons explicitly, and in section
\ref{effective operator} we derive in detail the effective transition
operator, which includes an anisotropic part of the form (1).  The
essential physical mechanism is that the electric field excites a
virtual phonon state $S_{tot}=0$ which is coupled to the $S_{tot}=1$
state by an anisotropic spin-phonon coupling which originates in
spin-orbit coupling.  An explanation involving the modulation of
static \DM interactions has been put forward recently for the case of
\navo.\cite{Gros} In that compound, however, no polarized experiments
are available and moreover, it is difficult to distinguish with a
magnetic dipole transition which turns out not to be forbidden by a
lattice selection rule.\cite{Sakai} The mechanism we develop here is
more general in that it does not require the presence of a static \DM
interaction. It only needs the \textit{instantaneous} breaking of the
inversion center which is assured by the appropriate phonons. This
allows us to consider the operator (\ref{effectiveoperator}) on the
strongest bonds irrespective of whether the bond lacks an inversion
center or not.  In section \ref{effective operator}, we give the
selection rules and the order of magnitude of such electric dipole
transitions. We compare with the experiments in $\rm CuGeO_3$ and $\rm
SrCu_2(BO_3)_2$ in section \ref{application}.

\section{Effective Magnetic Operator and Selection Rules}
\label{effective operator}

In this section we show that the first-order spin-orbit correction to
the spin-phonon coupling leads indeed to an effective magnetic
operator for the optical transitions.  We note that a phonon-assisted
optical transition is the usual explanation for the occurrence of the
singlet $S_{tot}=0$ bound states of two magnon states in the spectrum
of the high Tc's cuprates.\cite{Lorenzana} The spin-orbit correction
should then lead to transition to $S_{tot}=1$ states.

We start with a magnetic Hamiltonian for a chain or a layer of Cu atoms, for instance, that can be motivated by the usual super-exchange arguments:

\begin{equation}
H= \sum_{iad}  \textbf{\mbox{S}}_i \textbf{\mbox{J}}(\{\textbf{\mbox{u}}_{id}\}) \textbf{\mbox{S}}_{i+a} + H_{ph} - \textbf{\mbox{E}}.\textbf{\mbox{P}}_{ph}
\end{equation}
where $\textbf{\mbox{S}}_i$ is a spin operator,
$\textbf{\mbox{u}}_{id}$ is the displacement vector of the ion $d$ in
the unit-cell $i$, $H_{ph}$ is the phonon Hamiltonian which takes into
account the kinetic part of the ions and the spring constants,
$\textbf{\mbox{P}}_{ph}$ is the electric dipole of the ions and
$\textbf{\mbox{E}}$ is the external electric field. The magnetic
couplings, $\textbf{\mbox{J}}(\{\textbf{\mbox{u}}_{id}\})$, can be
expanded to first order in the ion displacements. Including the first
order in spin-orbit coupling, there is an extra term of \DM symmetry:

\begin{equation}
H_{sp} = \sum_{iad \alpha \beta} g_{d}^{\alpha} u^{\alpha}_{id}
 \textbf{\mbox{S}}_i. \textbf{\mbox{S}}_{i+a} + d^{\alpha \beta}_d u^{\alpha}_{id}
( \textbf{\mbox{S}}_i \times  \textbf{\mbox{S}}_{i+a})^{\beta} 
\label{spinphonon}
\end{equation}
where $g_d^{\alpha}$ is the partial derivative of the diagonal part of
$\textbf{\mbox{J}}(\{\textbf{\mbox{u}}_{id}\})$ with respect to
$\textbf{\mbox{u}}_{id}$ (it depends on the bond $i,a$ but we will not
write it explicitly in the following). The origin of $d^{\alpha
\beta}_d$ is explained below. This is indeed a general form for the
spin-phonon coupling and there is no restriction to be added on the
grounds of symmetry. The static \DM interaction is forbidden when
there is an inversion center at the middle of the bond. If the set of
displacements $\textbf{\mbox{u}}_{id}$ is such as to remove the
inversion center (which is the general case) then such an interaction
takes place. For example if we take the two symmetric ninety degrees
super-exchange paths Cu-O-Cu, there is a center of inversion and there
is an interference between the two paths that leads to no \DM
interaction. Suppose now that the two oxygens move upwards. Because
the hopping of the electrons is much faster than the typical phonon
frequency, the electrons see a frozen distorted lattice on that time
scale. The interference therefore does not occur anymore and there is
an effective \DM interaction linear in the displacements in the first
order. This is the origin of the second term of (\ref{spinphonon})
which involves a tensor $d_d^{\alpha \beta}$ since the displacements
in one direction, $\alpha$, generally produce a \DM vector in another
direction, $\beta$. Strictly speaking, $d_d^{\alpha \beta}$ also
depends upon the bond $i,a$, but we do not write it explicitly.  Note
that this term is derived in a super-exchange approach by taking into
account the spin-orbit coupling in first-order in perturbation theory
in the lines of the original Moriya's article.\cite{Moriya} We shall
refer to it as a \textit{dynamical} \DM interaction in the following.

The transition probability is then given at zero temperature
by the ``golden rule'':

\begin{eqnarray}
I(\omega) &=& \mid \langle f | \textbf{\mbox{E}}.\textbf{\mbox{P}}_{ph} | 0 \rangle \mid^2 \delta(\omega -
\omega_f) \\ \textbf{\mbox{P}}_{ph} &=& \sum_{id} q_d \textbf{\mbox{u}}_{id}
\label{transition dipolaire electrique}
\end{eqnarray}
\noindent
where $\omega_f$ is the energy of the excitation, typically the
one-magnon energy.  At first order in $H_{sp}$ in perturbation theory
the matrix element is written in terms of a sum over the excited
states:

\begin{eqnarray}
\langle  f | \textbf{\mbox{E}}.\textbf{\mbox{P}}_{ph} | 0 \rangle &=&
 \sum_{n} \frac{\langle f' |  \textbf{\mbox{E}}.\textbf{\mbox{P}}_{ph} | n \rangle \langle n | H_{sp} | 0'
\rangle}{\omega_0 - \omega_{n}} \nonumber \\
&+& \sum_{n} \frac{\langle f' | H_{sp} | n \rangle \langle
n | \textbf{\mbox{E}}.\textbf{\mbox{P}}_{ph} | 0' \rangle}{\omega_f - \omega_{n}}
\label{perturbation}
\end{eqnarray}
The intermediate states that contribute to the sum over $n$ contain
one phonon (whereas the initial and final states we are interested in do
not contain any phonon). The partial phonon matrix elements are
calculated out, but we keep the general form for the magnetic states
at this stage. In other words the phonons are integrated out and we end up with  an effective matrix element acting
between different magnetic states:

\begin{eqnarray}
\langle f |  \textbf{\mbox{E}}.\textbf{\mbox{P}}_{ph}| 0 \rangle &=& 
\langle f' \mid \sum_{ia} \gamma \textbf{\mbox{S}}_{i}. \textbf{\mbox{S}}_{i+a} +\mbox{\boldmath$\delta$} . ( \textbf{\mbox{S}}_{i} \times  \textbf{\mbox{S}}_{i+a}) \mid 0' \rangle \label{result} \\
\gamma &=&  \sum_s \frac{ \Omega_s}{\omega_f^2 - \Omega_s^2} g_s  (\textbf{\mbox{D}}_s. \textbf{\mbox{E}}) \\
\mbox{\boldmath$\delta$} &=&  \sum_s \frac{ \Omega_s}{\omega_f^2 - \Omega_s^2} \textbf{\mbox{d}}_s  (\textbf{\mbox{D}}_s. \textbf{\mbox{E}})
\label{delta}
\end{eqnarray}

\noindent 
where $\textbf{\mbox{D}}_s= \sum_d q_d
\mbox{\boldmath$\lambda$}_{dsq=0}$ is the amplitude of the
instantaneous electric dipole of the unit cell due to the phonon mode
$s$ with energy $\Omega_s=\Omega_{q=0,s}$. The final magnetic state
has an energy $\omega_f$. $g_s=\sum_{d,\alpha} g_d^{\alpha}
\lambda^{\alpha}_{ds}$ is the amplitude of the variation of the
magnetic exchange energy due the atomic distortions of the phonon $s$
($\lambda^{\alpha}_{ds}$ is the amplitude of the motion of the atom
$d$, in the direction $\alpha$ due to the phonon $s$ at $q=0$).
Similarly, $d_s^{\alpha}=\sum_{\beta d} d_d^{\alpha \beta}
\lambda^{\beta}_{ds}$ is the amplitude of the instantaneous
Dzyaloshinski-Moriya vector due to the phonon $s$.  The resulting
$\gamma$ and $\mbox{\boldmath$\delta$}$ depend on the bond considered.
They would usually couple the nearest neighbors, but could be
introduced for neighbors at larger distances if such super-exchange
processes were likely to take place. They can be introduced on the
basis of the symmetry which is usually reduced with respect to the
crystal symmetry by the presence of the external electric field.  Thus
we have written an effective operator announced in
eq. \ref{effectiveoperator} with $\textbf{\mbox{A}}_{\beta \gamma}=\frac{\partial \mbox{\boldmath$\delta$}^{\gamma}}{\partial \textbf{\mbox{E}}^{\beta}}$.

The selection rules are:

\begin{itemize}
\item (i)
$ \textbf{\mbox{D}}_s. \textbf{\mbox{E}} \neq 0$: the virtual phonon $s$ creates distortions that carry an instantaneous electric dipole $\textbf{\mbox{D}}_s$. In other words, the phonon $s$ must be infra-red active.
\item (ii) \begin{itemize}
\item $g_s \neq 0$: The distortion of the unit cell due to the phonon
$s$ modulates the magnetic exchange between the spins. The transition
at $\Delta S_{tot}=0$ is allowed.
\item $\textbf{\mbox{d}}_s \neq 0$: It implies that the distortion of the unit
cell due to the phonon $s$ must break instantaneously the symmetry by
inversion at the middle of the bond; so that to allow an instantaneous
Dzyaloshinski-Moriya interaction which amplitude is given by
$\textbf{\mbox{d}}_s$. The transitions between states that differ by the spin, $\Delta S_{tot}=1$,
are allowed and have an intensity $\sim \delta^2$.
\end{itemize}
\end{itemize}

Suppose that there is only one phonon mode $s$ which gives a major
contribution to the sum. In addition, we know that this active phonon
mode will appear in the infrared spectrum at the energy
$\Omega_{q=0,s}$, with an intensity given by
$I_{ph,s}=(\textbf{\mbox{D}}_s.\textbf{\mbox{E}})^2$. We can therefore
rewrite the intensity of the $\Delta S_{tot}=1$ line as:

\begin{equation}
I_e = \left[ \frac{1}{2} \frac{ \Omega_s  \textbf{\mbox{d}}_s}{\omega_f^2 - \Omega_s^2} \right]^2 I_{\mbox{\scriptsize{ph,s}}}
\label{probabilite dipolaire electrique estimation 1}
\end{equation} 
We denote by $E$ the order of magnitude of the variation of the
magnetic exchange energy due to the phonon and following Moriya,\cite{Moriya} we estimate $d_s \sim (\frac{\Delta g}{g}) E$. That
gives:

\begin{equation}
I_e \sim \left( \frac{\Delta g}{g} \right)^2 \left[ \frac{ \Omega_s E}{\omega_f^2 - \Omega_s^2} \right]^2 I_{\mbox{\scriptsize{ph,s}}}
\label{probabilite dipolaire electrique estimation 2}
\end{equation} 
This expression gives the intensity of such a process compared to the
intensity of the optically active phonon. It is reduced by two
factors: the spin-orbit coupling (in the cuprate materials, $\Delta
g/g$ can be $0.1$) and the ratio of the energy modulation of the
magnetic exchange due to the phonon by roughly the energy of the same
phonon. The latter is difficult to estimate : in $\rm CuGeO_3$, the
first optical phonons have $\Omega \sim 10 \rm meV$, and the
modulation can be as large as $E \sim 1 \rm meV$.\cite{Braden} That
gives $I_e \sim 10^{-4} I_{ph}$.

Another way to compare with is to consider that singlet excited
states, as for example the $S=0$ bound-state below the continuum in
$\rm CuGeO_3$, appear in the optical spectrum due to the isotropic
spin-phonon coupling (the $\gamma$ term). We denote their intensity by
$I_e^{singlet}$. Then we have: $I_e^{triplet} \sim \left( \frac{\Delta
g}{g} \right)^2 I_e^{singlet}$.  It means that if the singlet
bound-state appears in the optical spectrum with an intensity
$I_e^{singlet}$ due to the isotropic spin-phonon coupling, the triplet
states should also appear with an intensity which is roughly 100 times
smaller, if Moriya's estimate applies. 

\textit{Effect of a magnetic field}.  We consider a basic triplet
excitation here. A magnetic field lifts the degeneracy of the triplet
into three branches. When $\textbf{\mbox{H}} \parallel
\mbox{\boldmath$\delta$}$ ($\parallel z$), $S^z$ is a good quantum
number and the transition should satisfy $\Delta S^z=0$. Therefore, only
the mode $S^z=0$ could be observed and its intensity does not depend
on the strength of the field. By contrast, when the magnetic field is
perpendicular to $\mbox{\boldmath$\delta$}$, the wave-function is a
superposition of wave-functions with different $S^z$:

\begin{eqnarray}
 \Psi^{\pm \prime} &=& \frac{1}{\sqrt{2}} |1,0 \rangle + \nonumber \\
&\pm& \left( \frac{H_{\perp}}{2 \mid
 H_{\perp} \mid}  |1,1 \rangle + \frac{H_{\perp}^{*}}{2 \mid H_{\perp} \mid}
  |1,-1 \rangle \right) \\ \Psi^{0^{\prime}} &=& \frac{1}{\sqrt{2}}
 \left(\frac{H_{\perp}^{*}}{ \mid H_{\perp} \mid}  |1,-1 \rangle - \frac{H_{\perp}}{ \mid
 H_{\perp} \mid} |1,1 \rangle \right)
\end{eqnarray}
where the vector notation stands for $|S,S^z \rangle$.
The transition is allowed to the states $\Psi^{\pm \prime}$ with
quantum numbers $S^{\perp}=\pm 1$ and the mode which energy does not
depend on the field has no intensity. The magnetic field dependence is
therefore very different from what is expected for magnetic dipole
transitions.\cite{Sakai} This is basically because the
electric field conserves the $S^z$ quantum number. As we have
just seen, however, in transverse magnetic field, $S^z$ is no longer conserved
and the magnetic field-dependent branches may appear in the optical
spectrum. They do indeed appear  in \cugeo.\cite{Loosdrecht}

We now compare  the intensities of the
magnetic dipole transitions with those of the electric dipole
transitions that we have made explicit here. To make such a comparison, we consider the
following two models that give intensity to the optical
transitions. First a purely magnetic model and magnetic dipole
transitions. In order to have an intensity, we need to add a static
magnetic anisotropy, such as a Dzyaloshinski-Moriya interaction or an
anisotropy in the $g$ factor, which are both first order in the
spin-orbit coupling, so that in the most favourable  case (when no lattice
selection rule forbids it), the matrix element is of order $\sim
\Delta g/g$ at best.  In the second model, we consider an isotropic
magnetic model, but we add the anisotropic spin-phonon coupling that
we have considered above. The intensities of the
transitions of the two  models are given by:

\begin{eqnarray}
I_M &=& |\langle f | g\mu_B \textbf{\mbox{h}}.\textbf{\mbox{S}}_{tot} |
0 \rangle |^2 \sim [g \mu_B h]^2 \left( \frac{\Delta g}{g}
\right)^2 \\ I_E &=& \mid \langle f |
\textbf{\mbox{D}}. \textbf{\mbox{E}} \mid 0 \rangle \mid^2  \sim  \left( \frac{\Delta g}{g} \right)^2
 \left(\frac{ \Omega_s E}{\omega_f^2 -
\Omega_s^2}\right)^2 (\textbf{\mbox{D}}_s.\textbf{\mbox{E}})^2
\end{eqnarray}
So that the ratio is:

\begin{equation}
\frac{I_E}{I_M} \sim \left(\frac{\textbf{\mbox{D}}_s.\textbf{\mbox{E}}}{g \mu_B H} \right)^2  \left(\frac{ \Omega_s E}{\omega_f^2 - \Omega_s^2}\right)^2
\end{equation}

\noindent where $E=cH$, $c$ is the speed of light. $D$ is given by 
$D \sim e \lambda$ where $\lambda \sim
\sqrt{\frac{\hbar^2}{M \Omega}}$ is the amplitude of the motion of the ion and $e$ is its charge.

\begin{equation}
\frac{I_E}{I_M} \sim  \left( \frac{ec \lambda}{g \mu_B} \right)^2  \left(\frac{ \Omega E}{\omega_f^2 - \Omega^2}\right)^2 \sim 40 
\end{equation}

\noindent
with $M_{Cu}=63 \rm g/mol$ ($M_{at} \sim 10^{-25}\rm kg$), $\Omega =
10 \rm meV$, we find $\lambda \sim 0.1 \rm \AA$.  $g\mu_B=120 \mu \rm
eV/T$. We take $\omega=5 \rm meV$ for the energy of the magnetic mode
and $g=2 \rm meV$ for the spin-phonon coupling.  This estimation has
to be taken with a pinch of salt because of the the crude order of magnitude
given above, but it shows that there is no particular reason to not
consider the electric dipole transition due to dynamical
Dzyaloshinski-Moriya interaction.

\section{Application to  $\rm CuGeO_3$ and $\rm SrCu_2(BO_3)_2$.}
\label{application}

We compare the selection rules derived above with the
experimental observation in $\rm CuGeO_3$. Experimentally, the
absorption has been observed in the configuration $\textbf{\mbox{E}} \perp c$
but an extinction has been reported for $\textbf{\mbox{E}} \parallel c$,\cite{Damascelli} even in the presence of a magnetic field.\cite{Room-cugeo}

\begin{figure}[tbp]
\centerline{
\psfig{file=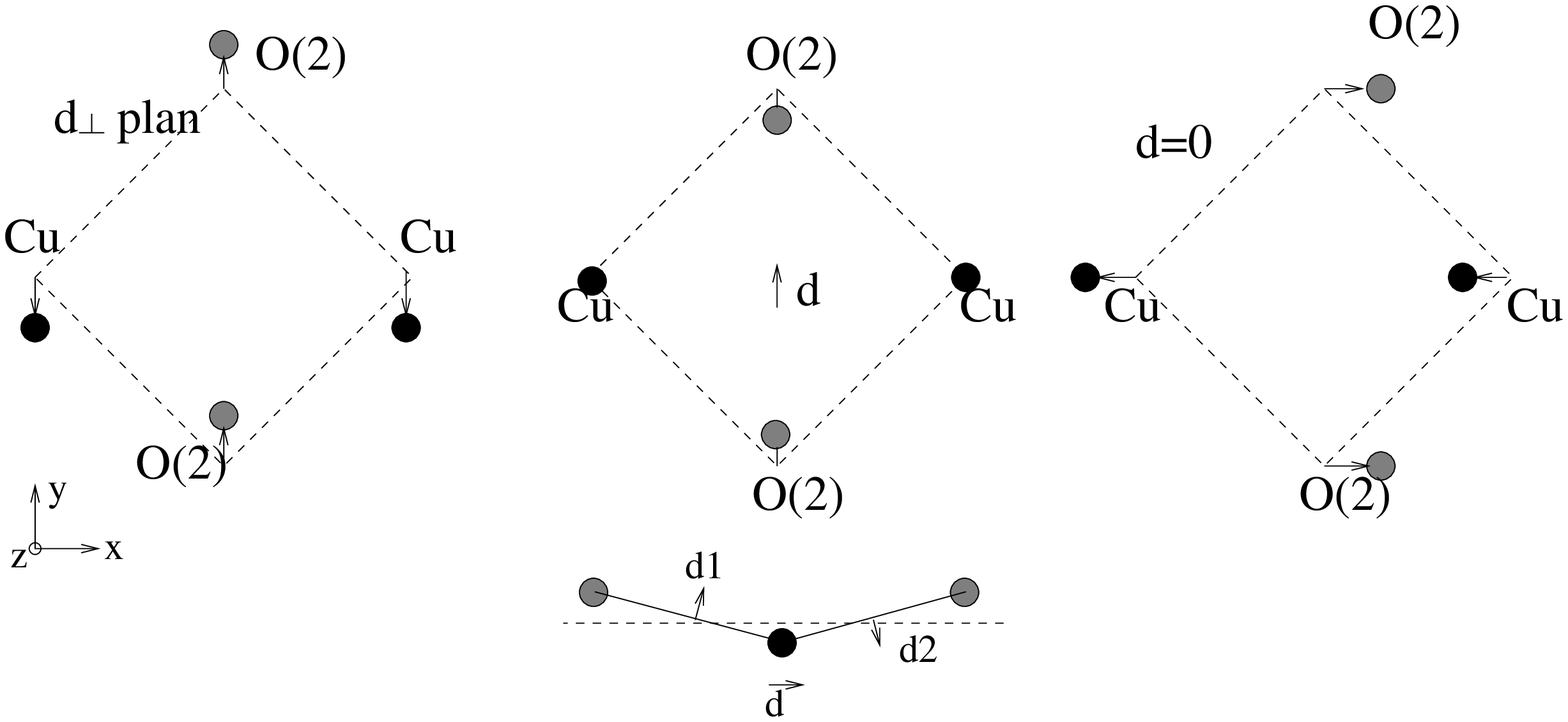,width=9cm,angle=0}}
\vspace{0.4cm}
\caption{Three examples of distortions of the $\rm Cu_2O_2$ cluster with associated Dzyaloshinski-Moriya vectors.
On the left, the motion along the $y$-axis creates a Dzyaloshinski-Moriya interaction whose vector is along $z$ (the mirror plane $\rm Cu_2O_2$ and the one perpendicular which has the O(2) atoms); on the middle, the atoms move out of the plane and the Dzyaloshinski-Moriya vector is along $y$ (mirror plane $xz$ passing through the bond Cu-Cu) ; on the right, these distortions break the inversion center at the middle of the $\rm Cu-Cu$ bond. However, the two perpendicular mirror planes $xy$ and $xz$ imply that the Dzyaloshinski-Moriya interaction actually vanishes. For CuGeO$_3$, the $x$-axis is the $c$-axis and the $xy$ plane is the plane of the CuO$_2$ chains.}
\label{d}
\end{figure}

We have a natural interpretation of this fact: when $\textbf{\mbox{E}}
\parallel c$, the only contributions to $\mbox{\boldmath$\delta$}$
come from the virtual phonons $s$ that have $\textbf{\mbox{D}}_s
\parallel c$, or, in other words, the virtual phonons involved are
those which are optically active in this configuration. The vector
$\mbox{\boldmath$\delta$}$ is given by $\sum_{\beta} d_d^{\alpha
\beta} \lambda_{ds,q=0}^{\beta}$ where $\lambda_{ds,q=0}^{\beta}$ are
the displacements of the atoms, the same as those that appear at
higher energy in the real phonon state $s$.  In the configuration
$\textbf{\mbox{E}} \parallel c$, the atoms in the phonon state $s$
roughly move along the $c$-axis. In a crystal with many atoms per unit
cell, this is not exactly true and the displacements will acquire
other components (a full study of the phonons that have been
theoretically predicted in Ref. \onlinecite{Braden3} does not change
the picture). Then, according to the figure \ref{d}, the dynamical
Dzyaloshinski-Moriya interaction is forbidden $\textbf{\mbox{d}}_s=0$
because of the mirror plane containing the atoms and the mirror plane
perpendicular to the previous plane and containing the Cu
atoms. Therefore, the intensity vanishes in this special
configuration. In other configurations, however, there is no such
symmetry arguments leading to a cancellation of the dynamical
Dzyaloshinski-Moriya interaction, and an intensity is expected in
agreement with the experiment performed in $\rm CuGeO_3$.

\medskip

\begin{figure}[tbp]
\centerline{
\psfig{file=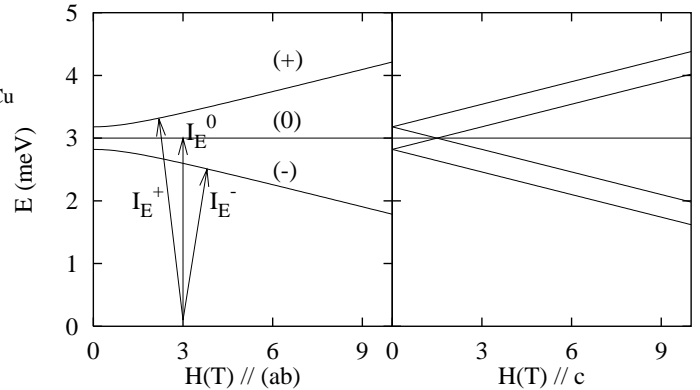,width=6cm,angle=-90}}
\vspace{0.2cm}
\caption{Excitation spectrum in \borate for two directions of the external magnetic field (from Ref. \protect \onlinecite{Cepas}). Definitions of the intensities of the optical transitions are also given.}
\label{excitationspectrum}
\end{figure}

We now consider the electric dipole transitions in $\rm
SrCu_2(BO_3)_2$ in greater detail. The obvious advantage of this
compound is that, neglecting anisotropies, it is described by the Shastry-Sutherland\cite{ShastrySutherland}  Hamiltonian that  possesses an exactly known ground state as a product of  local
singlets.\cite{Miyahara} Optical transitions have been observed
between this ground state and each of the zero-field three-split
triplet states \cite{Nojiri,Room} (see Fig. \ref{excitationspectrum})
that have been described previously.\cite{Cepas}  The probability of a
transition between the ground state $\Psi_0$ and an excited state $f$
is given by:

\begin{eqnarray}
\langle f |  \sum_{nn} \gamma \textbf{\mbox{S}}_i.\textbf{\mbox{S}}_j +
 \mbox{\boldmath$\delta$}_{ij}.(\textbf{\mbox{S}}_i \times \textbf{\mbox{S}}_j)
   | \Psi_0 \rangle
\label{general pattern0}
\end{eqnarray}
We have restricted the operator $H_E$ to the nearest neighbor spins
(nn) in order to find the largest effect.  The first part of it does
not change the total spin but may generate transitions to the first
excited states if the system has some anisotropy. We have considered
previously the existence of a Dzyaloshinski-Moriya interaction whose
vector is perpendicular to the plane.\cite{Cepas} We have shown that
such first-order anisotropy does not give intensity within the
assumption of magnetic dipole transitions. Here we start by
considering the electric dipole transitions generated by the first
part of the operator (\ref{general pattern0}) and in presence of the
static Dzyaloshinski-Moriya interaction. Using a symmetry argument we
show that this part actually vanishes. $S^z$ is a conserved quantity
so that we only need to consider the matrix elements with a $S^z=0$
final state.  The symmetry by the mirror plane perpendicular to the
(ab) plane and passing through a dimer is a symmetry of the
crystal. In this symmetry, the ground state and the operator
$\sum_{nn} \gamma \textbf{\mbox{S}}_i.\textbf{\mbox{S}}_j$ are both
even. However the triplet state $S^z=0$ adiabatically connected to the
local triplet at $J^{\prime}=0$ (the next nearest neighbor exchange)
is odd. Then the matrix element vanishes.  Additional spin anisotropy
of the \DM symmetry with extra in-plane components is present because
of the small buckling of the crystal structure at low
temperatures.\cite{Kakurai} However, this (together with possible
exchange anisotropies) would, in any case, respect the same
mirror-plane symmetry.  So the first term is not expected to give
intensity, because of this special symmetry.

In $\rm SrCu_2(BO_3)_2$, the transitions have been studied using
polarized electromagnetic waves and exhibit very peculiar polarisation
properties: in the configuration $\textbf{\mbox{E}} \parallel (ab)$,
at zero field, only the state at 24.2cm$^{-1}$ (i.e. the $S^z=0$ state
[the middle state]) appears in the spectrum, but an external in-plane
magnetic field gives intensity in the two other modes (upper and lower
modes).\cite{Room} Similarly when the magnetic field lies in the
$(ab)$ plane, only the upper state at 25.4cm$^{-1}$ (i.e. the $S^z=
\pm 1$) appears at zero magnetic field while an in-plane magnetic
field allows  observation of  the middle state, but not the lower one.
\cite{Room}

We now show that these observations are compatible with the dynamical
Dzyaloshinski-Moriya interaction which leads to the second part of the
effective operator (\ref{effectiveoperator}). To explain these results
we need to find the particular pattern of \textit{dynamical}
Dzyaloshinski-Moriya vectors and then the
$\mbox{\boldmath$\delta$}_{ij}$. That crucially depends on the
direction of the electric field of the wave, according to
eq. (\ref{delta}). In the following, we will determine the
$\mbox{\boldmath$\delta$}_{ij}$ but we restrict them to nearest
neighbor interactions.

\medskip

\textit{Configuration $\textbf{\mbox{E}}(t) \parallel (ab)$}.  Let us
consider first the case of a wave-vector of the electromagnetic wave
parallel to the $c$-axis, then the electric field lies in the $ab$
plane. According to the first selection rule (i), only the virtual
phonons which carry an electric dipole $D_s \parallel (ab)$ may
contribute to the sum (\ref{delta}). We basically assume that the main
displacements of the atoms in such a virtual phonon mode are confined
into the (ab) plane.   We make
the assumption that the main components of
$\mbox{\boldmath$\lambda$}_{ds}$ are parallel to the electric field,
so that we should be able to find the main components of the
Dzyaloshinski-Moriya vectors $\textbf{\mbox{d}}_{ij,s}$
(eq. \ref{delta}).  To estimate them (and then
$\mbox{\boldmath$\delta$}_{ij}$), we fix the atoms $d$ at the
distorted positions $\mbox{\boldmath$\lambda$}_{ds}$ and we then apply
the Moriya's rules which give the constraints on the
Dzyaloshinski-Moriya vectors.  In this case, the plane remains
instantaneously an approximate mirror plane for the crystal
structure. Subsequently, the instantaneous $d$-vector between the
spins, generated by the distortions, should be perpendicular to this
plane (parallel to the $c$-axis). The effective operator is therefore
written:

\begin{eqnarray}
H_{E \parallel (ab)} &=& \sum_{nn,A} \delta_z^A (\textbf{\mbox{S}}_i \times \textbf{\mbox{S}}_j)^z + \sum_{nn,B} \delta_{z}^B (\textbf{\mbox{S}}_i \times \textbf{\mbox{S}}_j)^z 
\label{confEab}
\end{eqnarray}
where $z$ is here again the $c$-axis. We have introduced two different
$\delta_z^{A,B}$ to take into account the existence of two dimers per
unit-cell. Taking the same would not change the argument. In the
following we take the notation
$\delta_z^2=[(\delta_z^{A})^2+(\delta_z^{B})^2]/2$.  The operator
(\ref{confEab}) does not break the symmetry by rotation around the
$c$-axis. A transition to the $S^z=\pm 1$ when the external magnetic
field is parallel to the $c$-axis is still forbidden. Only the $S^z=0$
triplet mode (at the middle of the others \cite{Cepas}) is allowed to
appear in the spectrum (this is in agreement with the general symmetry
argument given above since the electric field breaks the symmetry by
mirror plane).  This is in agreement with the experimental result at
zero-field.\cite{Room} We further predict that a magnetic field
parallel to the $c$-axis does not change the picture and gives no
intensity in the other branches. We can give an estimation of the
intensity assuming an approximate wave-function for the excited state
that we take from the strong dimerization limit.  In this
approximation, the excitation with $S^z=0$ is a purely local triplet
on the dimer A or B. This gives an intensity:

\begin{eqnarray}
I_E^0(H_{\parallel}) &=& |\langle \Psi_{q=0}^{A,0}|H_E|\Psi_0 \rangle|^2+|\langle \Psi_{q=0}^{B,0}|H_E|\Psi_0 \rangle|^2 = \delta_z^2/2 \\
I_E^{\pm}(H_{\parallel}) &=&  0 
\end{eqnarray}
We now consider the effect of a transverse magnetic field
($\textbf{\mbox{H}} \perp c$) on the intensities. A transverse
magnetic field splits the modes into three branches (figure
\ref{excitationspectrum}, left). To evaluate the intensity of each
branch, we first calculate the excited states in the approximation
used above, taking into account the \textit{static} \DM interaction
which is responsible for the zero-field splitting. Note that the other
in-plane components do not play any role in the triplet spectrum at
$q=0$,\cite{Kakurai} so that only the perpendicular component appear
in the following.

\begin{figure}[tbp]
\centerline{
\psfig{file=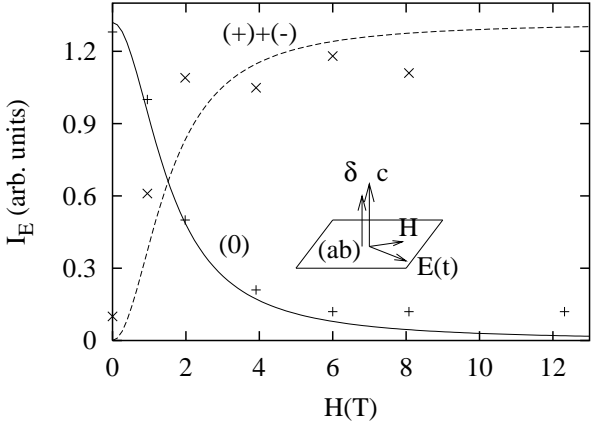,width=6cm,angle=-90}}
\vspace{0.2cm}
\caption{ Configuration $\textbf{\mbox{E}}(t) \parallel (ab)$. $\textbf{\mbox{H}} \parallel (ab)$. Intensity of the optical transitions from the ground state to the first split triplet state in \borate in the electric dipole approximation ($\mbox{\boldmath$\delta$} \parallel c$). $(+)+(-)$ is the sum of the intensity of the upper and lower mode  and $(0)$ is the middle mode. The theoretical curves are given by eqs. (\ref{ieparab1}) and (\ref{ieparab2}). There are no fitting parameters except an overall amplitude. The experimental data are from Ref. \protect \onlinecite{Room}. Note the intensity of the (+)+(-) mode is here twice as large as shown in the original experimental paper.\cite{noteRoom}}
\label{transitions optiques sous champ magnetique}
\end{figure}

The eigenvalues are in fact twice degenerate. The eigenvectors are
denoted by $\Psi_{q=0}^{(\pm,0)}$ and $\Psi_{q=0}^{(\pm,0) \prime}$
with energies $E_q^{(\pm,0)}$.  We then calculate the matrix elements
as a function of the transverse magnetic field:

\begin{eqnarray}
I_{E \parallel (ab)}^{(\pm,0)}(H_{\perp}) \equiv |\langle \Psi_{q=0}^{(\pm,0)}|H_E|\Psi_0 \rangle|^2+|\langle \Psi_{q=0}^{(\mp,0) \prime}|H_E|\Psi_0 \rangle|^2
\end{eqnarray}
We find:

\begin{eqnarray}
I^{0}_E(H_{\perp})   &=& \frac{\delta_z^2}{2} \frac{1}{1+h^2} \label{ieparab1}  \\ 
I^{\pm}_E(H_{\perp}) &=& \frac{\delta_z^2}{4} \frac{h^2}{1+h^2} \label{ieparab2}
\end{eqnarray} 
where $ h = g \mu_B H_{\perp}/2D $ is the transverse magnetic field in
the units of the \textit{static} Dzyaloshinski-Moriya interaction.  A
transverse field transfers intensity into the lower and upper modes.
The two curves given by $I^{0}_E(H_{\perp})$ and
$I^{+}_E(H_{\perp})+I^{-}_E(H_{\perp})$ are shown in figure
\ref{transitions optiques sous champ magnetique} together with the
experimental results of Ref. \onlinecite{Room}.  We have used the
non-renormalized value of $D=0.09 \rm meV$ extracted from the energy
spectrum \cite{Cepas} (all the calculations we performed here are in
the limit $J^{\prime}/J \rightarrow 0$, so that we use the value of
$D$ we would have extracted from such a calculation and not the
renormalized value).  Note that if we take $I^{0}_E(H_{\perp})$ and
$I^{+}_E(H_{\perp})$ for instance, they cross at a given field
$H_{\perp}=2\sqrt{2}D/(g\mu_B) \sim 2.1 T$, which is in good agreement
with the crossing of the fitted intensities in the original
experimental article ($H_{\perp}=2.3 T$)\cite{Room}. This is most
probably coincidental since we are using the wave-functions that are
not renormalized by the interaction $J^{\prime}$.

\medskip

\textit{Configuration $\textbf{\mbox{E}}(t) \parallel c$}.  We consider the
case of an electric field perpendicular to the plane $\textbf{\mbox{E}}(t)
\parallel c$. Let us suppose that the atoms move out of
plane. According to the figure \ref{d}, the dynamical
Dzyaloshinski-Moriya interaction would be in plane and perpendicular
to the $\rm Cu-Cu$ bond. The dimers are, however, perpendicular to one
another. Therefore the dynamical Dzyaloshinski-Moriya vectors of
adjacent dimers should be perpendicular as well. The effective
electric operator is:

\begin{equation}
H_{E \parallel c} = \sum_{nn,A} \mbox{\boldmath$\delta$}.(\textbf{\mbox{S}}_i \times \textbf{\mbox{S}}_j) + \sum_{nn,B}  \mbox{\boldmath$\delta$}^{\prime}.(\textbf{\mbox{S}}_i \times \textbf{\mbox{S}}_j) 
\end{equation}
where $\delta$ (respectively $\delta^{\prime}$) is perpendicular to the Cu-Cu
bond of the dimers $A$ (resp. B), so parallel to $y$ (resp. $x$). Note
that we take the same $|\delta|$ and $|\delta^{\prime}|$. Strictly
speaking there is no reason why they should be the same but taking
into account the special direction of the field we can reasonably
assume that the motions of the atoms which belong to adjacent dimers
are similar at least for the low-energy phonons. Let us apply this
operator on the ground state which is approximately a product of
singlet states on the dimers (we thus neglect the effect the static
Dzyaloshinski-Moriya interactions have on the ground state which would
give small corrections to the result). 

\begin{eqnarray}
H_{E \parallel c} |\Psi_0 \rangle  &=&  \frac{\delta}{2 \sqrt{2}} \left(\Psi_{q=0}^{A,S^z=+1} + i  \Psi_{q=0}^{B,S^z=+1} \right) \nonumber \\ &-&  \frac{\delta}{2 \sqrt{2}} \left( \Psi_{q=0}^{A,S^z=-1}  - i  \Psi_{q=0}^{B,S^z=-1} \right) \\
&=& \frac{ \delta}{2} \left( \Psi_{q=0}^{+,S^z=+1} -  \Psi_{q=0}^{-,S^z=-1} \right) 
\end{eqnarray}
Note that $\Psi_{q=0}^{+,S^z=+1}$ and $\Psi_{q=0}^{-,S^z=-1}$ are both
eigenstates of the Hamiltonian restricted to triplet states with the
same energy $J+2D$.  Depending on the sign of $D$, therefore, only the
\textit{upper} mode or the \textit{lower} mode should appear in the
spectrum. Experimentally, the upper mode has been found in such a
polarised configuration,\cite{Room} so that we conclude that
$D>0$. Only a detailed super-exchange calculation of $D$ would be able to infer it. The matrix elements giving the intensities are given by:

\begin{eqnarray}
I_{E \parallel c}^{+,+1}(H_{\parallel}) \equiv |\langle \Psi_{q=0}^{+,S^z=+1}|H_{E \parallel c}|\Psi_0 \rangle|^2= \delta^2/4 \\ I_{E \parallel c}^{-,-1}(H_{\parallel}) \equiv |\langle \Psi_{q=0}^{-,S^z=-1}|H_{E \parallel c}|\Psi_0 \rangle|^2= \delta^2/4
\end{eqnarray}
In zero external magnetic field, the two final states are degenerate
so that the total intensity of the optical transitions is the sum of
the two, i.e. $\delta^2/2$.  In a magnetic field parallel to the
c-axis ($z$-axis), the upper mode splits into two branches with equal
intensity $\delta^2/4$.

Furthermore, we calculate the intensities as a function of a
transverse magnetic field. The excited states $\Psi_{q=0}^{(\pm,0)}$
and $\Psi_{q=0}^{(\mp,0) \prime}$ are twice degenerate, so we
calculate:

\begin{eqnarray}
I_{E \parallel c}^{(\pm,0)}(H_{\perp}) \equiv |\langle \Psi_{q=0}^{(\pm,0)}|H_E|\Psi_0 \rangle|^2+|\langle \Psi_{q=0}^{(\mp,0) \prime}|H_E|\Psi_0 \rangle|^2
\end{eqnarray}
We find the following expressions for the intensity of the upper (+), lower (-) and  middle (0) states:

\begin{eqnarray}
I^{\pm}_{E \parallel c}(H_{\perp}) &=&  \frac{\delta^2}{8}  \frac{h^4}{\left[1 + h^2 \right] \left[\pm \sqrt{1+ h^2}-1\right]^2} \label{ieparc1} \\
I^{0}_{E \parallel c}(H_{\perp}) &=& \frac{\delta^2}{4} \frac{h^2}{1+h^2} 
\label{ieparc2}
\end{eqnarray}
where $ h = g \mu_B H_{\perp}/2D $.  The corresponding curves are
given in the figure \ref{transitions optiques sous champ
magnetique1}.  Note that the crossing between $I_E^+$ and $I_E^0$
occurs at $g \mu_B H_{\perp}=4 \sqrt{2} D$, therefore at a field two
times larger than in the configuration $\textbf{\mbox{E}} \parallel (ab)$.  The
agreement with the experiment is very good since such a balance of the
intensities has been observed.\cite{Room} The lower mode does not
actually appears in the spectrum experimentally and this is compatible
with the low intensity we found. If we take the non-renormalized value
of $D=0.09 \rm meV$, the crossing of the intensities occur at
$H_{\perp}=4.6 T$ which is in good agreement with the experimental
value ($\sim 6T$), as well as the overall behavior of the curves.

\begin{figure}[tbp]
\centerline{
\psfig{file=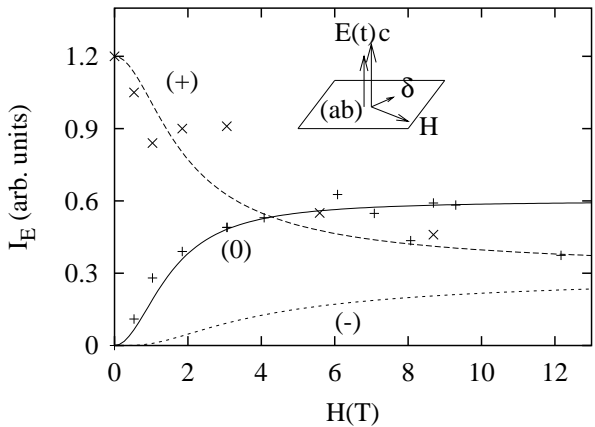,width=6cm,angle=-90}}
\vspace{0.2cm}
\caption{Configuration $\textbf{\mbox{E}}(t) \parallel c$. $\textbf{\mbox{H}} \parallel (ab)$. Intensity of the optical transitions from the ground state to the first split triplet state in \borate in the electric dipole approximation ($\mbox{\boldmath$\delta$} \perp c$). $(+),(-)$ and $(0)$ are, respectively, the upper, lower and middle mode. The theoretical curves are given by eqs. (\ref{ieparc1}) and (\ref{ieparc2}). There are no fitting parameters except an overall amplitude. The experimental data are from Ref. \protect \onlinecite{Room}.}
\label{transitions optiques sous champ magnetique1}
\end{figure}

\section{Conclusions}

In this paper, we have considered optical transitions with emission of
one magnetic excitation, $\Delta S_{tot}=1$. We give a mechanism in
terms of phonon-assisted transitions in which a virtual phonon is
involved. The selection rules of such processes were made explicit: in
brief we need a coupling to an infrared active phonon that breaks, at
least instantaneously, the symmetry of inversion between magnetically
coupled ions.  The intensity of such a process has been estimated and
we argue that it should be larger than a magnetic-dipole transition,
at least in systems in which spin-phonon couplings are appreciable.
It provides an alternative to purely electronic transitions that are
not allowed when an inversion center is present.

We note that we have considered uniquely the consequences of phonon
assisted optical transitions in the context of single-phonon
experiments, i.e. ESR and absorption. The same mechanism can
lead to processes in Raman scattering allowing single magnon creation,
with similar selection rules concerning centers of inversion in the
lattice. The effective operators will have similar symmetry but are
not identical, involving the polarisations of both incoming and
outgoing photons. Experimentally there are extra contributions linear
in both spin operators and spin-orbit couplings that are not present
in the single photon case. While for the spectroscopy of single
magnons in the materials studied, Raman scattering should be useful,
single photon experiments may permit more direct comparison with
microscopic estimates of intensities.

In the final section we have studied the two specific case of \cugeo
and \borate for which polarised experiments are available. We have
shown that predictions of the phonon-assisted theory agrees well both
with observed extinctions and also, for the case of \borate where
detailed results are available, with the dependence of intensities as
function of the external magnetic field. Further optical data should
be analysed in terms of an effective operator of the \DM symmetry for
the matrix elements in the electric dipole approximation.  Potentially
such optical experiments can provide a means of probing
microscopically the spin-phonon coupling which may be relevant to
other experiments, for example neutron inelastic scattering
experiments at finite momentum transfer, and a way of studying
four-spin correlation functions involving some sort of local
chiralities.

\medskip

We would like to thank T. R$\tilde{\mbox{o}}$$\tilde{\mbox{o}}$m for
correspondance and for providing us with his experimental results,
J.-P. Boucher, H. Nojiri, and T. Sakai for stimulating
discussions. O.C acknowledges financial supports from the
I.L.L. and the Indo-French grant IFCPAR/2404.1.

\end{document}